On Free Energy and Internal Combustion Engine Cycles.

William D Harris
548 43rd Street Apt A
Oakland, CA 94609
wdharris31416@gmail.com

Abstract: The performance of one type of Internal Combustion Engine (ICE) cycle is analyzed within the framework of thermodynamic free energies. ICE performance is different from that of an External Combustion Engine (ECE) which is dictated by Carnot's rule.

The changes in – respectively – the Gibbs Free Energy (G) and the Helmholtz Free Energy (A)

$$\Delta G = \Delta H - T\Delta S, \qquad (01)$$

$$\Delta A = \Delta E - T\Delta S, \qquad (02)$$

are quite familiar to scientists and engineers [1,2,3,4]. Attributes H, E, S, and T are respectively enthalpy, internal energy, entropy, and temperature. Relations (01) and (02) are restatements of the conservation of energy. For isobaric and isothermal processes, changes in energy ($\Delta H$) come into 2 forms: work ($\Delta G$) and heat ($T\Delta S$). Isochoric and isothermal processes follow relation (02).

There is some awkwardness in putting restrictions on the path taken by the process when G, A, and S are state functions, whose changes are path independent. Suppose one has a reactor whose input reactant flow is at the same pressure (P) and temperature (T) as the output product flow.

Let the heat exchange between the reactor and the surroundings take place at the same T. Also, all the processes within the reactor are done reversibly. Then the obtained work from the reactor is given by relation (01).

However, with the input and output streams remaining the same, if the reactor exchanges heat with the surroundings at temperature $T_{exch}$, say, different from aforementioned T, then the resulting work is not $\Delta G$ but some work defined as "free work" $W_{free}$:

$$W_{free} = \Delta H - T_{exch}\Delta S . \qquad (03)$$

$W_{free}$ is not constrained to cases where the initial (T,P) is the same as the output (T,P). $W_{free}$ can be generalized as

$$W_{free} = \Delta B - T_{exch}\Delta S . \qquad (04)$$

Depending on the process in question, $\Delta B$ can be $(H_2-H_1)$, $(E_2 - E_1)$, $(E_2-H_1)$, $(H_2-E_1)$, etc.. Subscripts 1 and 2 denote respectively the initial state of the input reactant and the final state of the output product. $\Delta S$ still remains $(S_2-S_1)$. Temperatures $T_1$, $T_2$, and $T_{exch}$ may be different from one another. Pressure $P_2$ may be different from $P_1$.

Given the temperature and pressure of the input reactant and of the output product, along with the heat exchange temperature $T_{exch}$, $W_{free}$ is the maximum work to be extracted. Often $W_{free}$ is called exergy [5]: "**The maximum fraction of an energy from which (in a reversible process) can be transformed into work is called exergy**". Thus relation (04) holds when all processes within the reactor are reversible. In practice, there is always some irreversibility denoted by the increased entropy $\Delta S_{irr}$ which has a nonzero and positive value. Thus maximum work $W_{free}$ is not achieved in practice and the actual work W obtained is

$$W = \Delta B - T_{exch}(\Delta S - \Delta S_{irr}). \tag{05}$$

An internal combustion engine (ICE) cycle with Carnot-like characteristics is studied within the "free" work framework. Previous studies [6,7,8] of ICE cycles from an exergy point of view looked at only the overall performance of the engine and not at the inside mechanisms of an ICE cycle. Relation (05) plays a key role since combustion is an irreversible process. For a Carnot-like engine, the non-adiabatic steps are all isothermal at some temperature. For simplicity's sake, let $T_1 = T_2 = T_{exch}$ and $P_2 = P_1$. The engine process is characterized as a flow process and thus $\Delta B = (H_2 - H_1)$. In relation (05), the path-dependent quantity is $\Delta S_{irr}$. Thus the focus of this analysis is on entropy changes. All one needs to know about the work obtained at each step is that the sum of the work for all the steps is equal to W in (05). The steps of the engine process are as follows.

Step 1. Reactant is inputted into the engine at $(P_1, T_1)$

  Net entropy change with respect to reactant at $(P_1, T_1) = 0$.

  Nothing has happened so far.

Step 2. Reactant is compressed adiabatically and reversibly to $(P_H, T_H)$

  Net entropy change with respect to reactant at $(P_1, T_1) = 0$.

  Adiabatic and reversible = isentropic.

Step 3. Reversible, isothermal, and isobaric energy extraction at $(P_H, T_H)$. The energy extraction can be, say, electrochemical. The reactant to product process is completed.

  Net entropy change with respect to reactant at $(P_1, T_1) = \Delta S_{rxn}(T_H, P_H)$, $\tag{06}$

  where $\Delta S_{rxn}(T_H, P_H)$ = entropy change of reaction at $(P_H, T_H)$.

  Electric energy extracted = $\Delta H_{rxn}(T_H, P_H) - T_H \Delta S_{rxn}(T_H, P_H)$, $\tag{07}$

  $= \Delta G_{rxn}(T_H, P_H)$. $\tag{08}$

  To maintain constant temperature for this step 3, the engine has an Internal Heat Reservoir (IHR) to absorb or provide thermal energy if $\Delta S_{rxn}(T_H, P_H)$ is negative or positive respectively.

Step 4. Isothermal reversible reversal of the heat flow between reactor and the IHR in step 3.

The final pressure is $P_M$ which is lower than $P_H$. Temperature is still at $T_H$.

Net entropy change with respect to reactant at $(P_1, T_1) = 0$.

(Step 3 + Step 4) = adiabatic and reversible = no entropy change.

At the end of step 4, the state of the IHR is the same as at the beginning of step 3.

Step 5. Isothermal dissipation of the electric energy obtained in Step 3 into thermal energy.

This step is needed to simulate the irreversible nature of combustion.

The final pressure is $P_N$ which is lower than $P_M$. Temperature is still at $T_H$.

Net $\Delta S$ with respect to reactant at $(P_1, T_1) = \Delta S_{irr}$

$$= \Delta S_{rxn}(T_H, P_H) - \Delta H_{rxn}(T_H, P_H)/T_H, \qquad (09)$$

$$= -\Delta G_{rxn}(T_H, P_H)/T_H. \qquad (10)$$

The combustion reaction is spontaneous, thus $\Delta G < 0$. Relation (10) shows that $\Delta S_{irr} > 0$.

Step 6. Reversible and adiabatic expansion of the product from $(P_N, T_H)$ to $(P_Q, T_1)$.

Net $\Delta S$ with respect to reactant at $(P_1, T_1) = -\Delta G_{rxn}(T_H, P_H)/T_H$.

Step 6, being reversible and adiabatic, does not change the net entropy change.

Step 7. Reversible cooling/heating of product from $(P_Q, T_1)$ to $(P_1, T_1)$.

Net $\Delta S$ with respect to reactant at $(P_1, T_1) = \Delta S_{rxn}(T_1, P_1)$. $\qquad (11)$

Relation (11) is just the definition of entropy change of reaction at constant T and P.

Total transfer of thermal energy between the ICE and the surroundings is

$$- T_1 \Delta S_{rxn}(T_1, P_1) - (T_1/T_H) \Delta G_{rxn}(T_H, P_H) \qquad (12)$$

Overall (from step 1 to step 7), the total energy involved is

$$\Delta B = (H_2 - H_1) = \Delta H_{rxn}(T_1, P_1). \tag{13}$$

Combining relations (05), (12), and (13) gives the net work W:

$$W = \Delta H_{rxn}(T_1, P_1) - T_1 \Delta S_{rxn}(T_1, P_1) - (T_1/T_H)\Delta G_{rxn}(T_H, P_H). \tag{14}$$

Using relations (07) and (08), one rewrites W as

$$W = \Delta G_{rxn}(T_1, P_1) - (T_1/T_H)\Delta G_{rxn}(T_H, P_H). \tag{15}$$

Relation (15) is the main and central result of this note and is different from the formula of the work obtained from a Carnot Heat engine:

$$W = \Delta H_H [1 - (T_1/T_H)], \tag{16}$$

the subscripts 1 and H denote respectively the initial temperature and the combustion temperature in current model of a Carnot-like ICE. Relation (16) has been used by automotive engineers for many years with success, although the common engines follow the Otto or Diesel cycles. To better understand the differences between (15) and (16), relation (15) is expanded:

$$W = [\Delta H(T_1, P_1) - (T_1/T_H)\Delta H(T_H, P_H)] - T_1[\Delta S(T_1, P_1) - \Delta S(T_H, P_H)]. \tag{17}$$

Usually in combustion processes, the T$\Delta$S terms are about 20 times smaller than the $\Delta H$ terms. In addition, relation (17) involves taking the differences in the entropy changes. Thus the second term in (17), depending explicitly on the various $\Delta S$'s, can be neglected without great error for the common experimental conditions of automotive engine performance. Besides, the oxidant is air which is about 80% nitrogen, a mostly spectator gas, thus the second term in (17) decreases in magnitude further more.

Under the common experimental conditions of automotive engine performance, the enthalpy change of combustion may be safely assumed to change little with temperature. To understand the last statement, the differential change in enthalpy is given in terms of differentials in temperature (dT) and in pressure (dP):

$$dH = C_P dT - T(\partial V/\partial T)_P dP + V dP , \qquad (18)$$

Where $C_P$ is the heat capacity at constant pressure and $(\partial V/\partial T)_P$ is the thermal expansion factor at constant pressure. Under the same experimental conditions, the reactants and products act mostly like ideal gases and the enthalpy differential becomes:

$$dH \approx \Delta C_P dT , \qquad (19)$$

Where $\Delta C_P$ is the difference in the constant pressure heat capacities between reactant and product. Often one finds that

$$\left|\Delta C_P (T_H - T_l)\right| \ll \left|\Delta H_{rxn}\right| , \qquad (20)$$

and that relation (16) is a good approximation of the work obtained from the above described Carnot-like ICE.

Relation (16) gives the exact work obtained from a Carnot External Combustion Engine (ECE). The entropy change from the external combustion has no influence on the behavior of the working fluid inside the engine. In addition, for the ECE, the working fluid either maintains a constant chemical composition or undergoes a cyclical change in chemical composition. However for an ICE, there is no cyclical composition change since the input is reactant and the output is product and there is not transition from product to reactant. For an ICE, the entropy change of the combustion reaction must play a role in the engine thermodynamic performance.

The role of ΔH as source of potential work is different for ICE's and ECE's. For ECE's, all of

ΔH is potential source for work. For ICE's, for the ($\Delta S_{rxn} < 0$) case, only part of ΔH is the potential source of work. Keeping in mind that ΔH = ΔG +TΔS, for ICE's, $\Delta G_{rxn}$ is the potential source of work. $T\Delta S_{rxn}$ is intimately connected to the fact that product was produced from reactant: once all of $\Delta G_{rxn}$ is converted to work, any attempt to convert $T\Delta S_{rxn}$ into work will fail and bring about the synthesis of reactant from product, a process that does not occur in ICE's. In the ($\Delta S_{rxn} > 0$) case, ΔH is the potential source of most but not all work: heat inflow into the ICE is needed for complete conversion into work.

Let us revisit relation (16). Assuming that $\Delta G_{rxn}$ remains finite at all temperatures, then

$$\text{as } T_H \to \infty, \; W \to \Delta G_{rxn}(T_1, P_1). \tag{21}$$

There is no reason not to assume that $\Delta G_{rxn}$ remains finite for all temperatures. At relative low temperatures, there are differences in the thermodynamics properties of reactant and product and these differences can be such to increase the magnitude of $\Delta G_{rxn}$ as temperature increases. However, well before infinite temperature is reached, both reactant and product will dissociate totally and act like ideal gases and $\Delta G_{rxn}$ will approach some asymptotic value. Focusing on relation (17), one finds that

$$\text{as } T_H \to \infty, \; \Delta S_{rxn}(T_H) \to 0. \tag{22}$$

The pressure has no role in the limiting behavior – with respect to infinite temperature -of the entropy change of reaction and is thus not stated explicitly in relation (22).

Relation (22) implies an intriguing non-chemical method of determining if compound X can be produced from or produce compound Y using only Pressure, Volume, Temperature and Calorimetric data. The method is as follows:

Step A- Find the entropy of X and Y at ($P_1, T_1$). The Zeroth (sometimes called Third) law of

Thermodynamics states that entropies of all compounds at 0K is zero. The (P,V,T) and calorimetric data are used to find the starting entropies.

Step B- Keep the entropy X at constant entropy by compressing adiabatically (and reversibly) the compound to high pressure and temperature ($P_H$, $T_H$).

Step C- Bring compound Y to the same ($P_H$, $T_H$) by a series of adiabatic compression and subsequent (and necessary) cooling/heating. Of course infinite temperature cannot be actually reached (besides, all this analysis excludes nuclear reactions). Nonetheless, for every high temperature reached by compound X, one brings compound Y to the same T and P as compound X and record the entropy change in Y.

Step D. One plots the entropy change between X and Y for the set of increasing high temperatures. If the extrapolated entropy difference between X and Y is still appreciably non-zero then there is no chemical connection between X and Y.

At very high temperature, all compounds eventually dissociates into a plasma state. What seems to matter is the elemental composition of the compounds. If compound X is composed of, say, carbon and hydrogen and compound Y is composed of, say, oxygen and nitrogen, then one should not expect the entropies of the plasmas to be the same. There may be some ratios of oxygen and nitrogen which has the same asymptotic entropy value as for a given (carbon, hydrogen) ratio. Said cases are happenstance and in no way imply that X can be produced by or produce Y.

Let us revisit the issue of irreversibility. Irreversible paths from thermodynamic state A to thermodynamic state B are much more numerous than the number of reversible paths. There is at least another model for a Carnot-type ICE, where step 3 is now the reversible, isothermal, and isochoric energy extraction at ($V_H$, $T_H$), $V_H$ is the volume of the compressed reactant at the end of the reversible adiabatic compression (step 2). Steps 4, and 5 are modified such that the temperature remains at $T_H$. Now $\Delta S_{irr}$ becomes $-\Delta A_{rxn}(T_H,V_H)/T_H$, A being the Helmholtz free energy, and - via application of relation (05) - the effective work output is

$$W = \Delta G_{rxn}(T_1,P_1) - (T_1/T_H)\Delta A_{rxn}(T_H,V_H) . \qquad (23)$$

In practice, W may range between relations (15) and (23). The take-away lesson is that for ICE, the heat of combustion, $Q_H$, is not the total potential source of work but some effective free energy $\Delta \overline{K}_{rxn}(T_H,P_H) \equiv (Q_H - T_H \Delta \overline{S}_{rxn})$, where $\Delta \overline{S}_{rxn}$ is the effective entropy change of combustion. Relations (15) and (23) can be generalized to

$$W = \Delta G_{rxn}(T_1,P_1) - (T_1/T_H)\Delta \overline{K}_{rxn}(T_H,V_H) . \qquad (24)$$

One can do a similar analysis for the other ICE cycles such as Otto and Diesel. However, please note that there is no unique and precise formula for W. Given that $\Delta G_{rxn}$ is a function of T and P, for the Carnot-like ICE, one can have Step 3 to consume just part of the reactant, proceed with Steps 4 and 5, return to Step 3 to consume more of the reactant, and repeat the cycle until all the reactant is consumed. Then the work W in relations (14,15,17) will have different values.

The closing point is that the performance of an ICE is different from that of an ECE and the basis of this difference is the role of the entropy change of combustion plays in the ICE dynamics. Hydrogen can be used in an electrochemical fuel cell or in an ICE. Relation (24) provides one equation which can describe both fuel cells and ICE's.